**Author:** Eugenio Petrovich

**Title:** Accumulation of Knowledge in Para-Scientific Areas. The Case of Analytic Philosophy

**Affiliation and Address:** Department of Philosophy, University of Milan. Via Festa del Perdono 7, 20122 Milan (Italy)

**E-mail address:** eugenio.petrovich@unimi.it

**Telephone number:** 0039 331 48 79 302

**ORCID ID:** 0000-0001-9646-0471

**Acknowledgments:** My gratitude goes first to Ludo Waltman (CWTS Leiden), who was my supervisor during my visiting period at CWTS, for very helpful advice and precious insights. I would like to thank also my advisor Luca Guzzardi and my colleagues in the Doctoral School in Philosophy and Human Sciences at the University of Milan (especially Valerio Buonomo, Daniele Cassaghi and Emiliano Tolusso) for discussing with me several parts of this research. I would finally like to thank two anonymous reviewers for providing useful suggestions.

# Accumulation of Knowledge in Para-Scientific Areas. The Case of Analytic Philosophy

**Eugenio Petrovich**[1]

**Abstract:** This study analyzes how the accumulation of knowledge takes place in para-scientific areas, focusing on the case of Analytic Philosophy. The theoretical framework chosen for the analysis is Kuhn's theory of normal science. The methodology employed is qualitative citation context analysis. A sample of 60 papers published in leading Analytic Philosophy journals between 1950 and 2009 is analyzed, and a specific classificatory scheme is developed to classify citations according to their epistemological function. Compared to previous studies of citation context, this is the first paper that includes the temporal dimension into the analysis of citation context, in order to gain insights into the process of knowledge accumulation. Interestingly, the results show that Analytic Philosophy started accumulating after Second World War, but in a peculiar way. The accumulation was not matched by a corresponding rising consensus. This can be explained by the hypothesis that AP underwent a process of fragmentation in sub-fields during the second half of the century.

## 1 INTRODUCTION

One of the most striking features of science is the progressive accumulation of scientific contents in time. The stock of scientific knowledge available is indeed not fixed but continues to grow, and, especially in the Twentieth century, with an exponential pace (Price 1986). Even if the idea that science progresses in a linear and cumulative manner has been challenged since the landmark work of Thomas Kuhn by different philosophers, historians and sociologists of science, it is undeniable that *scientific information* grows.[2] And this growth is surely one of the components of the very idea of "scientific

[1] Department of Philosophy, University of Milan. Via Festa del Perdono 7, 20122 Milan (Italy)

[2] In this study, no explicit definition of the term "knowledge" will be provided, since any definition is bound to raise difficult epistemological and, more broadly, philosophical problems. However, this study inscribes itself within an approach to knowledge that can be termed "anti-representationalism" (Hacking 1979). Anti-representationalism is rooted in Kant and Hegel's philosophy of knowledge and at least some of its features have been embraced in the Twentieth century by Popper, Lakatos, Kuhn, and Foucault (see Popper 1979, Lakatos 1978, Kuhn 1970 and Foucault 2002). Anti-representationalism considers knowledge as the *product* of the *activity* of the *knowing subject* (be it the individual or a collective entity such as the scientific community), not as a "true representation" of the "world out there" or as a set of "justified true beliefs" held by knowing subjects. According to anti-representationalism, the primary feature of knowledge is its being a (human) product. As a product, knowledge constitutes a "third world" (different from both the "first world", the physical reality, and the "second world", the mental reality) that is materially embodied in books, archives, databases and scientific papers (Popper 1979 : ch. 3). From this point of view, saying that scientific knowledge grows is the same as saying that *scientific literature* grows. A full justification of anti-representationalism lies outside the scope of this paper. However, the main motivations for adopting it as a theoretical background are the following. First, it allows to avoid vast philosophical debates about truth and

progress" (Hacking 1979). Kuhn himself recognized such dimension, identifying it within what he called "normal science" (i.e. the science practiced between scientific revolutions, see Kuhn 1970).

If the process of accumulation taking place in scientific disciplines is widely studied by sociologists of science and scientometricians, the way in which information accumulates and knowledge grows in non-scientific areas is less known. However, it is important to study such cases, as they shed light on similarities and differences in the way sciences and humanities produce knowledge. In this regard, philosophy is an extremely interesting case-study. Philosophy is a venerable discipline, with a two-thousand-year-old history. Even if today it is usually classified as a humanistic field, it shares with science the interest for developing theories in a rational and rigorous way. In particular, contemporary Analytic Philosophy seems to be attracted by a scientific model of inquiry, mimicking several features of contemporary science in its research practices and self-image.

This study focuses on Analytic Philosophy as a case-study of *para-science* and uses citation analysis to investigate how knowledge accumulates within it. The aim is to assess if and how the pattern of accumulation taking place in Analytic Philosophy has become analogous to the one sciences display.

The paper is organized as follows. Firstly, in the remainder of the Introduction, Analytic Philosophy is presented in detail and its alleged similarities with science are discussed. Secondly, a definite research question is pointed out in light of the Kuhnian theory of normal science. Thirdly, the general method which was designed to address the research question (qualitative citation context analysis) is introduced, and related literature is reviewed. The following section (Methodology) describes in detail how the citation context analysis was performed, highlighting the features of the sample, the considered time frames, and the classificatory scheme designed to label citations. In the Result section, the main empirical results of the analysis are presented, and in the Discussion section, they are discussed in light of the research question and the wider theoretical framework under investigation (i.e. the accumulation process of Analytic Philosophy and science). Lastly, in the Conclusions section, the most important findings are highlighted, and further lines of research are suggested.

### 1.1 PRESENTATION OF THE CASE-STUDY

Since the Second World War, Analytic Philosophy (hereafter AP) developed into the dominant philosophical tradition of the English-speaking world (UK, US, and Australia) and Scandinavian countries (Searle 1996). In the last two decades, it has attracted increasing attention also in Continental Europe, as indicated by the establishment of several societies of analytic philosophers all over Europe (Beaney 2013). Today, AP is one of the main traditions of twentieth-century Western philosophy, together with the "Continental Philosophy". The scission between these two traditions (the so-called "Analytic-Continental divide") is a key feature of the contemporary philosophical landscape. Even if both sides have made several attempts to bridge the gap, the split remains a clear sociological fact of contemporary philosophy (Levy 2003, Glock 2008).

However, from a historiographical point of view neither analytic philosophers nor historians of philosophy reached a consensus on a proper definition of Analytic Philosophy or, for that matters, Continental Philosophy. Glock (2008) reviews and carefully criticizes all the proposals made in the literature to assess distinctive features of AP, ranging from geo-linguistic definitions (AP as Anglo-

---

the definition of knowledge as "justified true belief" (an introduction to them can be found in Moran 2010 : ch. 11). Second, it allows to focus on the phenomenon of accumulation of knowledge without raising the problem of the advancement towards the "truth" (see Bird 2007 for a recent discussion of the relationship between scientific progress and truth), that is particularly difficult in the case of philosophical knowledge. Finally, it allows to inquire the *activity* of accumulation of knowledge and track its changes over time, without assuming a teleological drive such as "truth" in the history of knowledge (Kuhn 1970 : ch. 3 and Bloor 1991).

American philosophy) to doctrinal and topic criteria (lack of historical awareness, a special concern with language, the scarce attention for ethical and political issues), to more refined approaches, that focused on stylistic and methodological definitions (e.g. clarity in the philosophical argumentation, the method of analysis)

Glock concludes that none of the proposed criteria is immune to counter-examples. Thus, he advances the idea that what binds AP together is a Wittgensteinian "family resemblance", i.e. a thread of overlapping similarities (doctrinal, methodological and stylistic), along with a common historical origin, traceable back to the works of Gottlob Frege, Bertrand Russell, logical positivism, Willard V.O. Quine, Ludwig Wittgenstein and Oxford ordinary language philosophy (J. L. Austin, G. Ryle etc.).

### 1.2 A HYPOTHESIS ABOUT AP

Recently, both philosophers and historians of philosophy suggested the hypothesis that AP could be seen as a *para-science,* namely as an area of philosophy that underwent, during the twentieth century, a process of *progressive approaching* to the *methods*, the *ethos* and the *self-image* of the sciences.

Levy (2003) explicitly suggests that the research practice of AP can be successfully understood when compared to the Kuhnian normal science. He proposes to characterize AP as "philosophical research conducted under a paradigm":

> My suggestion is this: AP has successfully modeled itself on the physical sciences. Work in it is thus guided by paradigms that function in the way Kuhn sketches, and the discipline is reproduced in something akin to the way in which the sciences are reproduced. (Levy 2003 : 291)

Richardson (2008) notes that the adoption of a "scientific ethos" in philosophy, somehow resembling the Mertonian normative model of science (Merton 1973), was actively promoted by logical positivists (one of the roots of contemporary AP) already in the 1930s. Logical positivists advocated a piecemeal and collaborative style of work in philosophy and even created the word "philosophical research" on the model of "scientific research" (Reichenbach 1978, quoted in Richardson 2008).

Marconi (2014) and Putnam (1997), respectively leading analytic philosophers in Italy and in the USA, claim that the "the self-image of analytic philosophy is scientific rather than humanistic" (Putnam 1997 : 201), and that analytic philosophers today conceive AP as a collective enterprise pursued by highly specialized "professionals" (Marconi 2014), working in accordance with the best scientific practices (such as peer-review, publication in specialized journals, use of technical language, a high division of cognitive labor, and so on).

### 1.3 AIM AND RESEARCH QUESTION

The aim of this study is to empirically assess the claim that AP has undergone a process of approaching a scientific style of intellectual production. Specifically, this study will test the hypothesis, advanced by Levy (2003), that AP has undergone a process of "normalization", i.e. an assimilation to the Kuhnian normal science model.

In order to translate this general aim into a definite research question, it is essential to highlight what features of Kuhn's theory of normal science are taken into consideration for the purpose of this study. As it is well known, Kuhn (1970) closely tied the notion of normal science to the one of paradigm or, later, to what he called a "disciplinary matrix" (i.e. the set of symbolic generalization, metaphysical commitments, validation standards and scientific exemplars shared by a discipline). However, the notion of normal science in Kuhn (1970) is introduced as the stage following either the pre-paradigmatic proto-science or the revolutionary science, which is practiced during scientific revolutions. In this respect, what distinguishes normal science from the other stages is the fact that *only in normal science periods* scientific knowledge *accumulates*, i.e. develops in the classic cumulative way we usually attribute to sciences. During pre-paradigmatic or revolutionary science, scientific knowledge is instead unstable and continuously contested. From a sociological point of view, it can be said that in periods of normal science the *consensus* within the scientific community is high, whereas in pre-paradigmatic or revolutionary science the consensus is low, and the scientific community is fragmented in competing, incompatible schools (Cole 1992).

In the present study, we focus on the *cumulative nature* of normal science. Thus, we aim to answer the following research question:

> Has AP started to *accumulate* during the second half of the twentieth century?

It is important to untangle the notion of *accumulation* from the close, albeit different, notion of *progress towards the truth*. The accumulation is a process that concerns the growth of the available stock of knowledge, and it is a property of the knowledge itself. The *epistemological* question about the *relation* between the stock of knowledge and the "world out there", i.e. the question of the progress towards the truth, is a different issue, which is not addressed in the present study. [3]

## 1.4 CITATIONS AS AN "OPEN WINDOW" ON THE PROCESS OF ACCUMULATION

The method chosen to answer the research question is *citation context analysis*. To justify this choice, first, it will be explained why citation analysis was carried out, and second, why the context of citation was also taken into account.

It is well known that citations, both in sciences and in humanities, represent links among documents. These links incorporate a *diachronic* information: citations go (usually) from newer documents to older documents. [4] Citations are therefore one of the ways in which an author relates her own contribution to the existing stock of knowledge (Hyland 1999). In both sciences and humanities, this stock of knowledge, i.e. the disciplinary body, is often referred to as the "literature" of the field. In certain disciplines, the cited literature is confined to the very recent "research front" (Price 1970), whereas in others it may go back many years, sometimes even centuries (e.g. in Continental Philosophy, it is not unusual to cite Aristotle for theoretical purposes, treating him as a "contemporary" philosopher). Hence, citations are an open window on the process of accumulation, which is the focus of this study. Clearly, citations provide an incomplete perspective on the way in which knowledge accumulates. For instance, the phenomenon of OBI (obliteration by inclusion) pointed out by Merton (1988) is a case in which a certain piece of knowledge is incorporated into the disciplinary body without the use of explicit citation. However, citations are probably the most apparent and the most empirically accessible trace of the accumulation process. This is the reason why citation analysis was chosen as the method for the present research.

The links the citing document establishes with the cited documents (i.e. the citations) can be further characterized taking into account the *context* in which the citation appears, i.e. the portion of text that surrounds the citation. This approach to the study of citations is known as *citation context analysis* (Small 1982, Bornmann and Daniel 2008). Citation context analysis seems particularly suitable to clarify the fine-grained structure of the knowledge accumulation process since it allows to characterize the relationship between the citing and the cited text and consider the *epistemic* relations among documents (e.g. criticism, support, acknowledgement, etc.).

---

[3] Kuhn was very clear in distinguishing the accumulation process taking place in the normal science periods from the metaphysical notion of progress, conceived as a movement of approach to the truth. Indeed, he was quite wary about the very idea that science progresses towards the "truth", because this seems to imply a *teleological drive* in the development of science – a claim that is difficult to test empirically (See Kuhn 1970: ch. 13 and Bloor 1991).

[4] Rarely, citations point out also to the "future", in the case of references to "forthcoming" publications.

Citation context analysis has been performed either *quantitatively* or *qualitatively*. The quantitative approach (which is the most recent one) carries out the analysis of the context of citation by using natural language processing techniques that are trained based on an annotated corpus (e.g. Valenzuela et al. 2015, Bertin et al. 2016, Sula and Miller 2014, Catalini et al. 2015, Teufel et al. 2006). On the other hand, the qualitative approach adopts a classic close reading method, and experts are required to read and characterize the whole sentence (or even the whole paper) incorporating citations. In this study, the qualitative approach was chosen to carry out the citation context analysis. According to the excellent review by Bornmann and Daniel (2008), qualitative citation context studies can be divided in two families: a) studies focusing on the functions that citations play in the citing document (supporting claims, overviewing the field, etc.); and b) studies focusing on how a set of documents (e.g. the works by funding figures of a disciplines) is cited in the following literature.

Since this study belongs to the first family, the most relevant literature in this strand of research will be briefly reviewed. Moravcsick and Murugesan (1975) investigated citations in 30 articles in theoretical high energy physics and devised a classificatory scheme that considered five functions of the citations. The first (conceptual versus operational citations) and the third function (evolutionary vs. juxtapositional citations) were meant to provide insight into the type of connectedness of scientific communication, whereas the second (organic versus perfunctory) and the fourth dimension (confirmative versus negational citations) addressed directly the quality of the citations. The fifth dimension (valuable versus redundant) was related to the importance of the cited work for the citing work. The study of Moravcsick and Murugesan aimed to assess the use of citation scores as measures of scientific quality. According to the authors, the high percentage of perfunctory citations found in the papers (41%) casts "serious doubts" about the use of citations as an indication of quality. Chubin and Moitra (1975) followed up on Moravcsick and Murugesan's work and also studied citations in physics. They examined citations in letters and articles in major physics journals and devised a classificatory scheme which focused on defining citations as either affirmative or negative. They found that citations made by physicists were most frequently affirmative citations, whereas negational citations represented only a small fraction. The finding that scientists in general tend to avoid negative citations was corroborated by further studies (Catalini et al. 2015, Bertin et al. 2016, Cano 1989). Spiegel-Rösing (1977) is the first study of citation context outside the field of physics. Spiegel-Rösing examined citations in the first four volumes of the journal *Science Studies* and found that the most frequent function of references is to substantiate a statement made in the citing text or point out to further information. In fact, supportive citations represent 80% of the total. Frost (1979) is the first study to investigate the function of citations in a humanities area, i.e. German literary research. Frost discusses the differences as well as the similarities in the citation usage of sciences and humanities, highlighting the fact that in the humanities there is a clear difference between references pointing to primary literature and references to secondary literature (i.e. research produced by other scholars, see also Hellqvist 2009). The classificatory scheme developed by Frost will be discussed in paragraph 2.3, when the scheme used in this study will be presented. Finally, Peritz (1983) proposed a general scheme to classify citation roles (designed mainly for social sciences), which includes the following 8 categories of citations: setting the stage citations, citations providing background information, methodological, comparative, speculative, documentary, historical, and casual citations.

In the 1980s, researchers began to shift their focus to the study of how some specific document or set of documents was used by the following literature that cited it (see e.g. Oppenheim and Renn 1978, Hooten 1991). At the same time, new methodologies were used to investigate the citation behavior of researchers, most notably surveys and interviews (see sections 2.4 of Bornmann and Daniel 2008 for a review of these studies). In the last decade, quantitative approaches to citation context analysis, aiming to automatically recognize citations functions and quality, have considerably flourished (see Hernãndez-Alvarez and Gomez (2016) for a recent overview of the state of the art).

In scientometric and sociological literature, citation context analysis is sometimes used as an empirical tool to assess the main competing theories of citations, i.e. the normative theory inspired by Merton (e.g. Kaplan 1965) versus the socio-constructivist approach (e.g. Gilbert 1977) rooted in the social constructivist sociology of science. The idea is to test the predictions about citation behavior made by the two theories and check if they match real citation practices, as they emerge from the citation context. However, the focus of the present study is slightly different. The primary interest is not to test citation theories against the citation behavior of analytic philosophers, but to use citation context analysis to unveil the pattern of accumulation of knowledge in AP. Thus, the purpose of this study is more epistemological than sociological, in so far as it focuses primarily on an epistemological topic (the process of knowledge accumulation), rather than on the sociology of citation practices within an epistemic community.[5]

### 1.5 CITATION ANALYSIS OF PHILOSOPHY

To conclude the review of the literature, it is worth mentioning previous studies that investigated philosophy via diverse scientometric and citation analysis techniques.

Cullars (1998), Hyland (1999) and Sula and Miller (2014) are the closest to the present research. The first study examined 539 references from 183 English-language philosophy monographs published in 1984. Several characteristics of the citations were analyzed, including the language of the cited document, the gender of both citing and cited authors, the citing author's attitude towards the cited material, and the subject correlation between citing and cited documents. The author concluded that citation patterns in philosophy are typically humanistic, with the bulk of citations pointing to books rather than journal articles. Hyland (1999) explored the ways in which academic citation practices contribute to the construction of disciplinary knowledge. The author analyzed a multi-disciplinary corpus of 80 research articles from different disciplines (including philosophy) and interviewed experienced researchers about their citation practices. Hyland concluded that philosophy, in line with other soft disciplines such as marketing and sociology, employed more citations than hard-science disciplines such as physics and engineering. However, the author suggests that citation in philosophy plays a different role to the one it plays in the hard sciences: citations are not primarily used to extend the thread of knowledge but to position the writer in relation to views that the author supports or opposes. This is clear also from the choice of report verbs used by philosophers to introduce the citation: contrary to the sciences, where neutral report verbs (such as "present" and "report") are common, in philosophy evaluative report verbs (such as "overlook", "exaggerate"), that flag agreement or disagreement with the cited reference, are the most frequently used. Finally, Sula and Miller (2014) used sentiment analysis to automatically classify citations in four different humanistic journals (including *The Journal of Philosophy*) and discovered that philosophy showed the most negative polarity, i.e. philosophers frequently dissociate from cited documents, providing critical commentaries on others' work.

Other studies do not address philosophy in general but focus on specific sub-disciplines or philosophical topics. Kreuzman (2011), Wray (2010), Wray and Bornmann (2015), and Byron (2007) addressed philosophy of science. Kreuzman (2011) used co-citation analysis techniques to map the interaction between epistemology and philosophy of science. Byron (2007) focused on philosophy of biology, using bibliometric data to assess the traditional historical account of the emergence of this sub-discipline in philosophy of science. Brad Wray (2010) determined the key journals of philosophy of science by counting how many times those journals are cited in leading philosophy of science companions. Brad Wray (2015) used an advanced bibliometric method (RPYS, Referenced Publication Year Spectroscopy) to identify peaks in citations in philosophy of science.

[5] Clearly, this does not exclude the possibility that this research could be of interest for sociologists of science.

Ahlgren et al. (2015) used various science mapping techniques (co-citation analysis and co-occurrence of terms) to map topical subdomains in philosophy (free will and *sorites* debates). Analytic philosophy as a field was specifically addressed in Buonomo and Petrovich (2018). In this study, a set of 4966 papers published in top analytic philosophy journals was analyzed with VOSviewer (Van Eck and Waltman 2010). Lists of most cited authors were extracted and the process of specialization of AP was mapped by co-citation mapping. Specialization in philosophy (not only analytic) was the subject also of Wray (2014), who examined the degree of specialization in various sub-fields of philosophy, starting from data obtained from surveying professional philosophers.

Finally, it is worth mentioning Cronin et al. (2003), who chronicled the use of acknowledgements in Twentieth-century philosophy and psychology. They demonstrated how acknowledgement has gradually established itself as a constitutive element of academic writing, and how it can be used to reveal "subautorship" forms of collaboration.

# 2 METHODOLOGY

## 2.1 FROM INTELLECTUAL CONTENT TO PUBLICATIONS

The first step of this study was to translate the vague and debated notion of AP into a relevant set of publications upon which citation context analysis could be conducted. In order to avoid the historiographical issues mentioned in the Introduction section, it was decided to consider AP as a *research specialty* inside the wider discipline of philosophy. A research specialty can be defined as a

> self-organized network of researchers who tend to study the same research topics, attend the same conferences, read and cite each other's research papers and publish in the same research journals. (Morris et Van der Veer Martens 2008 : 214-215)

Defining AP as a research specialty clearly does not solve the historiographical problems stated above, but it allows to "operationalize" the vague notion of AP, and climb down from the level of the intellectual content to the level of the *publications,* which are the outcome of the intellectual activity. In this way, AP can be translated into a set of publications, following a familiar epistemological move of scientometric research (i.e. the reduction of the intellectual content of science to the public form of its outcome, namely publications, see Wouters 1999).

Papers in specialized journals were selected as the publication type to study, whereas books were not taken into account. This is mainly because of two reasons. First, there is already a study concerning citation context in philosophy monographs (Cullars 1998). Even if it does not focus on AP but on philosophy in general, it was chosen not to replicate this study but to focus on the journal literature (as Cullars himself suggests in the final section of the paper). Second, AP opts for journal articles rather than books as key outlets for disseminating research (Levy 2003, Alghren et al. 2015). [6]

Five journals were selected as representative of AP: *The Philosophical Review, Noûs, The Journal of Philosophy, Mind, Philosophy and Phenomenological Research*. These journals are all considered highly prestigious in the AP community. They were ranked the "top five" journals in a recent survey[7] conducted by the blog *Leiter Reports: A Philosophy Blog*, a popular blog among analytic philosophers. Furthermore, they are considered "generalist" journals, hence they are a good representation of the whole AP field.[8]

[6] This is another feature of AP that suggests an approaching to the normal science style of intellectual production.

### 2.2 The time-window and the sample

Since the research question studied in this paper is historical in nature, a wide time-window was chosen: from 1950 to 2009. Covering the last 60 years after the Second World War, this time-window allowed to detect the changes in the citation behavior of analytic philosophers, disclosing the patterns of the accumulation process.

The 60-year time-window was further divided into six 10-year timespans. For each of the six timespans, the list of the most cited papers, published in the selected journals in that decade, was obtained from Web of Science.[9] Then, for each of the six timespans, the 10 most cited papers were chosen, and their full-texts downloaded. The result was a sample of 60 papers, divided into six groups of 10 papers each.[10] Taken together, the total number of cited references in the sample amounted to 1293 references.

It is important to stress that this is the first study that considers the *temporal dimension* in the analysis of the citation context. In fact, previous studies (Moravcsik and Murugesan 1975, Chubin and Moitra 1975, Spiegel-Rösing 1977, Frost 1979, Bertin et al. 2016, Sula and Miller 2014, Catalini et al. 2015, Teufel et al. 2006) focused on disciplines whose citation practices were assumed to be rather "stable" in time, therefore no temporal dimension was considered. In the case of AP, however, it would be wrong not to take into consideration the temporal dimension. On the one hand, historians of philosophy agree on an evolution of AP in the last century: AP started from a small group of philosophical schools located in Vienna, Cambridge, Oxford, and a few American universities, and later on became a worldwide enterprise with thousands of practitioners (Kuklick 2001, Beaney 2013, Marconi 2014). It is unlikely that the citation behavior remained unchanged during this evolution. On the other hand, the accumulation of knowledge, as well as the approaching to Normal science, is structurally a *process,* i.e. something developing in time, not a *state*, i.e. something occurring in a definite moment. Therefore, it can be successfully detected only if the temporal dimension is included in the analysis.

Instead of a random sample, the top ten most cited papers for each decade were selected for the following reasons. First, the high citation score of these papers means that they were widely read (i.e. they had a great impact on the AP community). Second, many "classics" of AP do appear in the sample, confirming that the selected papers are a good representation of "high-quality" AP in each decade, as analytic philosophers asked to assess the list confirmed. Being the publications both widely cited and high quality, it may be argued that they play the role of Kuhnian paradigms in the AP community, setting, to a certain extent, standards of citation behavior.[11] Third, the choice of a random set of papers would have implied a sample of at least 100 papers for each decade, in order to obtain statistically significant results. The analysis of such a large number of papers, however, was practically impossible because of the close reading approach chosen for this study (see above).

[7] http://leiterreports.typepad.com/blog/2015/09/the-top-20-general-philosophy-journals-2015.html

[8] Buonomo and Petrovich (2018) discuss in detail how to select journals representative of mainstream and "high-quality" research in AP

[9] Search date: 29.08.2017

[10] Two papers originally ranked top ten were excluded and substituted with subsequent papers in the ranking: [Anderson 1958] and [Sen 1985]. Even if they were published in philosophy journals, their subject falls outside AP (even in the broad sense), being respectively a result in formal logic for the first and economic welfare theory for the second.

[11] See Chen (2013 : ch. 6) for some arguments in favor of the assimilation of high-impact publications to Kuhnian paradigms.

## 2.3 THE CLASSIFICATORY SCHEME

Once the sample was chosen, the next step was developing a *classificatory scheme* to classify the citations. Classificatory schemes previously proposed in the literature were not suitable to address the epistemological focus of this study or the intellectual features of a philosophical specialty like AP. With regard to the first point, studies such as Moravcsik and Murugesan (1975) and Chubin and Moitra (1975) aimed to test sociological theories of citations and developed classificatory schemes suited to that purpose. As regards the second point, classificatory schemes should be adapted to the epistemic practices of the discipline under study. For example, in the sciences there are citations whose function is to identify specific laboratory techniques or equipment (Garfield 1962). This does not occur in AP, where no laboratory activities are involved. The classificatory scheme should take into account such differences.

The classificatory scheme used in this study is summarized in Tab. 1. Six categories were defined (plus a seventh "Unknown" category designed to collect citations that did not fall in any of the categories). All the categories were designed to capture different aspects of the phenomenon of accumulation. In the Appendix 2 examples of each category are provided.

The category of "State of the art citations" captures citations that are employed to provide an overview of the field to which the citing paper intends to contribute. From an epistemological point of view, they point out the presence of a shared body of literature. The authors use such body as a *knowledge base* to articulate their own contribution around. In a mature field, the lack of such citations may result in the rejection of the paper, because the author did not sufficiently review the state of the art. The authors' attitude towards this kind of citations is "neutral": they neither endorse nor criticize them. The authors rather use the "State of the art citation" to locate their contribution in a specific stream of philosophical debates. The presence of these citations in AP is particularly interesting since they demonstrate that the AP research itself is divided into specific sub-contexts.

The "Supporting citations" category gathers the citations that support the argument proposed in the citing paper, either by supplying arguments for certain claims or by reinforcing the author's argument by showing that the cited reference agrees with it. From an epistemological point of view, they underline a *constructive* operation: knowledge is accumulated via a positive relation of agreement between the citing and the cited document.

The third category, "Supplementary/perfunctory citations", refers to citations that seem not to be essential to follow the author's main argument. [12] On the other hand, the fourth category, "Acknowledgement citations", is meant to capture citations that are explicitly used to pay homage to or acknowledge the cited author. [13] It may be argued that these categories are more sociological than epistemological, and thus not fitting the epistemological purpose of this study. However, it is important to emphasize that, even if these citations have no clear epistemological functions, they do provide *negative* or *indirect* epistemological information. If Supplementary/perfunctory citations were the most common type, it would mean that AP's accumulation process is only loosely connected with epistemological factors. Alternatively, if Acknowledgement citations resulted to be the most common type (perhaps a more plausible scenario), it could be argued that the AP's accumulation process is mainly based on social relations rather than epistemic. Both scenarios would offer important epistemological, *although negative*, insights into the accumulation process of AP: they would support the idea that AP accumulates in a "non-epistemic" manner. This would prompt further research in the social structure of AP. Hence, the more sociological categories, namely "Supplementary/perfunctory citation" and "Acknowledgement citations", were included in the present study.

---

[12] The further question about the reasons an author may have to cite supplementary material (e.g. hidden social-networking purposes) is not pursued in this study.

[13] In this study, only explicit citations were considered, i.e. citations pointing to specific documents. Proto-citations such as "I owe this point to Prof. X", commonly used in acknowledgements section of recent papers, are not counted. For an interesting study of acknowledgements in philosophy, see the mentioned study conducted by Cronin et al. (2003).

In the "Critical citations" category, the cited reference is criticized by the author, who marks her disagreement with the cited reference. In the sciences, this type of citation is quite uncommon (Catalini et al. 2015, Bertin et al. 2016, Cano 1989), but in AP it is reasonable to expect a significant number of these citations, because of the high value of dialectics in philosophy in general (Cullars 1998). From an epistemological point of view, a critical citation is a *destructive* operation: its function is to undermine a previous piece of knowledge. From a sociological point of view, it flags lack of consensus on a topic.

Finally, the last category, "Documental citation", covers citations to documents used as historical sources. The cited text is mentioned as a support for the historical reconstruction provided in the citing paper. This kind of citation is very common in humanities, where there is a clear distinction between citations pointing to the textual material (e.g. historical documents in history) and citations pointing to other scholars (Hellqvist 2009, Frost 1979). In a theoretical discipline like AP, the number of historical citations is expected to be quite low.

Frost (1979) proposes a classificatory scheme that partially aligns with the one used in the present study. Frost's first group of citations (type A, documentation of primary sources) corresponds with what here is called Documental citations; her second group (type B.1, «references to previous scholarship […] independent of approval or disapproval of the citing author», 406) matches the State of the art citations, and her third (B.2) and fourth (B.3) groups can be compared with the Supporting and Critical citations, respectively. However, the scheme here proposed differs from Frost's one as it includes categories that aim to detect also sociological aspects of the citation behavior, i.e. the categories Supplementary/perfunctory citations and Acknowledgement citations, that she did not consider.

| # | Type | Function |
|---|---|---|
| 1 | State of the art citation | The reference is used to provide an overview of the state of the art of the field. It is «neutral»: the citing author does not use it either to support her argument or to criticize the cited document. It includes citations to standard mental experiments or examples. |
| 2 | Supporting citation | The reference supports the argument of the citing author, either because it brings additional arguments to the stated claims or because it strengthens the author's arguments since the authors agree on specific topics. |
| 3 | Supplementary/perfunctory citation | The reference to the cited document is accidental. The cited work it is not essential to follow or support the citing author's argument. |
| 4 | Acknowledgment citation | The citing author pays a form of tribute or acknowledgement towards the cited document. |
| 5 | Critical citation | The cited document is criticized, the citing author disagrees with the cited document. |
| 6 | Documental citation | The cited document is used to support the historical reconstruction of the argument of the cited author provided by the citing author. |
| 7 | Unknown | The citation does not fit into any of the previous categories. |

**Table 1** Classificatory scheme

### 2.4 THE DECADE PROFILES

Each paper was read entirely, and every cited reference was assigned to one or more categories, often in light of the entire content of the paper.

When a reference was assigned to more than one category, its score was fractionalized in order to avoid percentages higher than 100% in the final sum.[14] The score of each reference was not weighted: documents that were mentioned multiple times in the citing paper did not receive a score proportional to the number of mentions. Equally, the score of documents which were cited both positively and critically was not fractionalized proportionally to the number of times in which they were cited either positively or critically.[15] The use of a system of weights, even if theoretically desirable, is actually unmanageable in practice. In fact, if adopted, it would have raised exegetical issues on each of the analyzed references, and it would have been ultimately highly subjective and not reproducible.

The results of the analysis of each paper were summarized in a paper profile, reporting for each category both the absolute score and the percentage on the total. Then, the profiles of the papers of the same timespan were aggregated to obtain a single profile for each of the six decades.

## 3 RESULTS

### 3.1 AGGREGATED LEVEL (1950-2009)

The results of the citation context analysis for the whole period (1950-2009) are summarized in Tab. 2. Data show that the Supporting citation is the most common type of citation (37.9%), followed by Critical citations (23.1%) and State of the art citations (21.4%). These three categories alone cover 82.4% of the citations. Supplementary, Acknowledgement and Documental citations play only a minor role on the total (overall, they account only for 11.4% of the citations).

| Rank | Type | Percentage on the total |
|---|---|---|
| 1 | Supporting citations | 37.9% |
| 2 | Critical citations | 23.1% |
| 3 | State of the art citations | 21.4% |
| 4 | Supplementary/perfunctory citations | 4.7% |
| 5 | Acknowledgment citations | 3.4% |
| 6 | Documental citations | 3.4% |
| 7 | Unknown | 1.1% |
| | *TOTAL (3 papers with no citatations)*[16] | *95.0%* |

**Table 2** Types of citation (1950-2009)

However, the data show a great variance in time, as the next section highlights and explains in detail.

[14] It turned out that no citation could be attributed to more than three categories.

[15] For example, if a reference was mentioned 3 times in the same article, but it played always the same function, it scored 1 point, and not 3 points. On the other hand, if it was mentioned 3 times, each time with a different function, its point was equally divided among the functions it played.

[16] 3 papers (1 in the decade 1950-1959 and 2 in the decade 1960-1969, 5% on the total) had no citations, so that the total is slightly lower than 100%.

### 3.2 DEVELOPMENT IN TIME

The first clear pattern is the increase in the average number of references per paper (Fig.1), which shows a five-fold increase from the 1950s (8.8 citations per paper) to 2000s (44.1 citations per paper). The increase between the 1980s and 1990s is particularly evident. Moreover, only in the first two decades, we find papers without citations (1 in the 1950s and 2 in the 1960s).

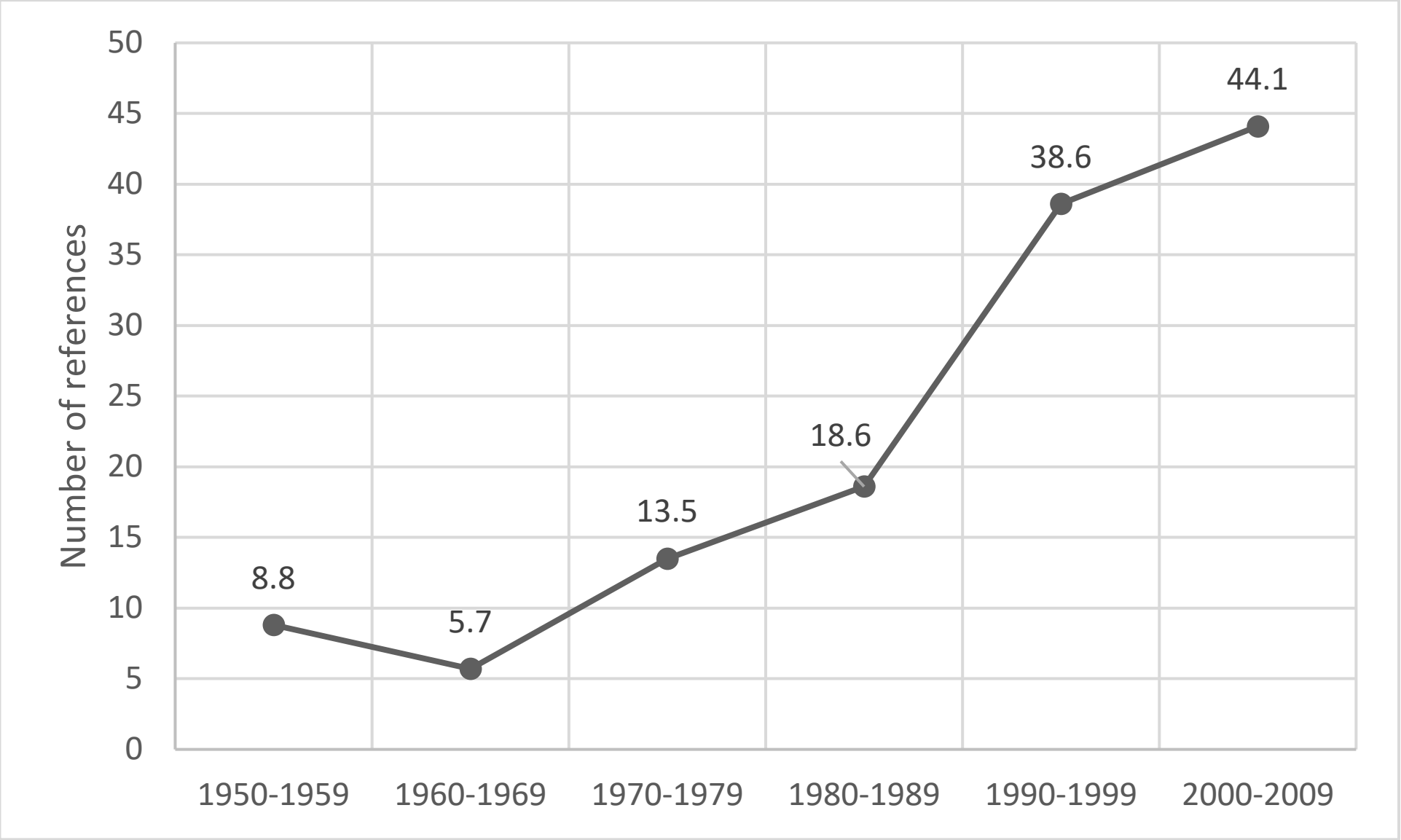

**Fig. 1** Average number of references per paper over time

Fig. 2 shows the average number of Supporting, Critical and State of the Art citations per paper in each decade. These three categories account for most of the citations in all the six timespans, and they all increase in time. This was expected, given the increase in the average number of reference shown in Fig.1. However, their increase rate is significantly different: State of the Art citations raise almost exponentially (increasing 13-folds, from 1.5 to 20.1 from the 1950s to 2000s), whereas Supporting citations increase quite linearly (from 4.0 to 13.4), and Critical citations slightly increase at first (from 1.5 to 4.4), but then even seem to decrease in the last decade.

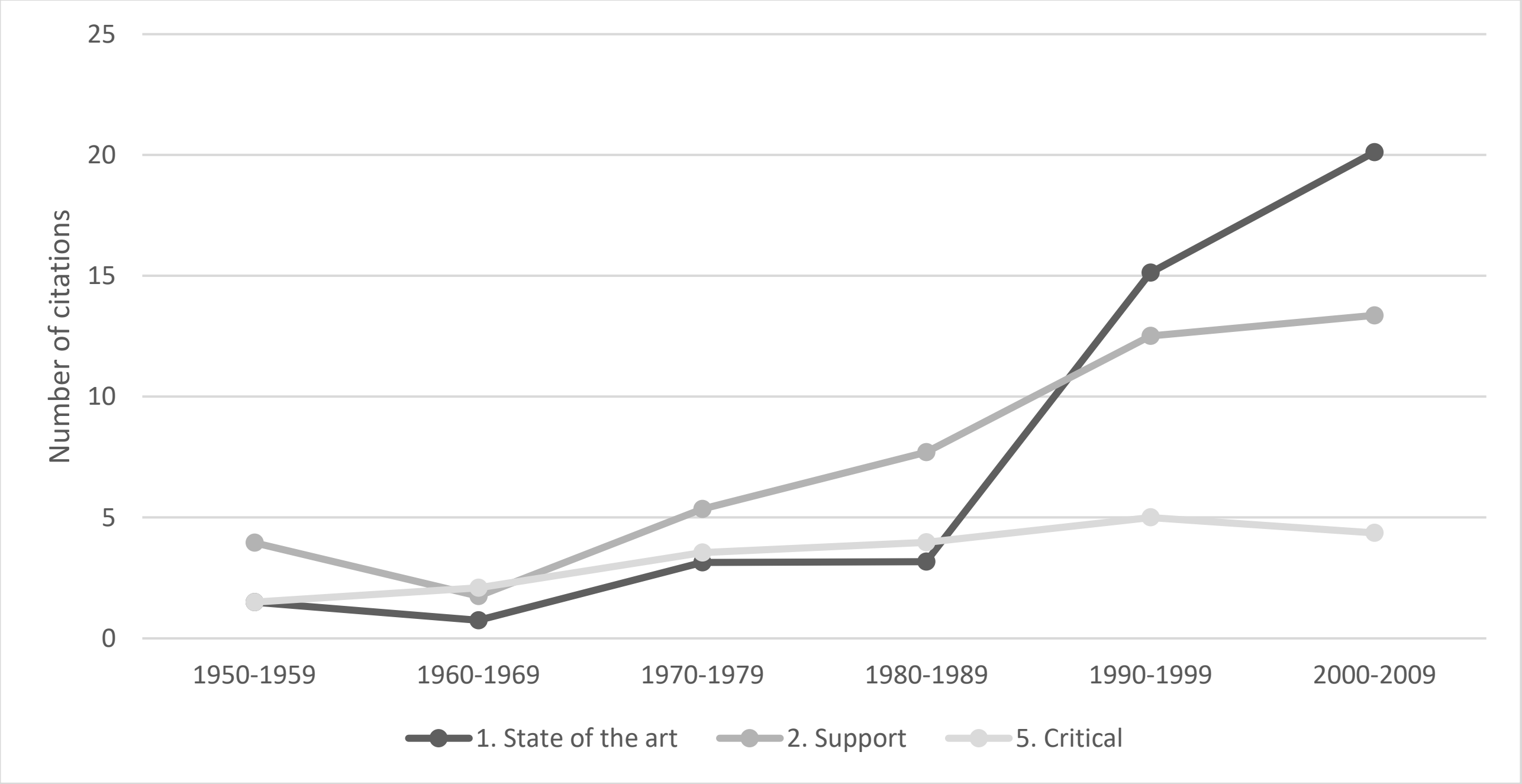

**Fig. 2** Average number of State of the art, Supporting and Critical citations per paper over time

In order to better understand how the relative weight of each category changes over time, we have considered the percentage of each citations category on the total.

Fig. 3, 4 and 5 show the distribution of Supporting, Critical and State of the Art citations in time by box plot diagrams. Again, these three categories account for most of the citations in all the six timespans.

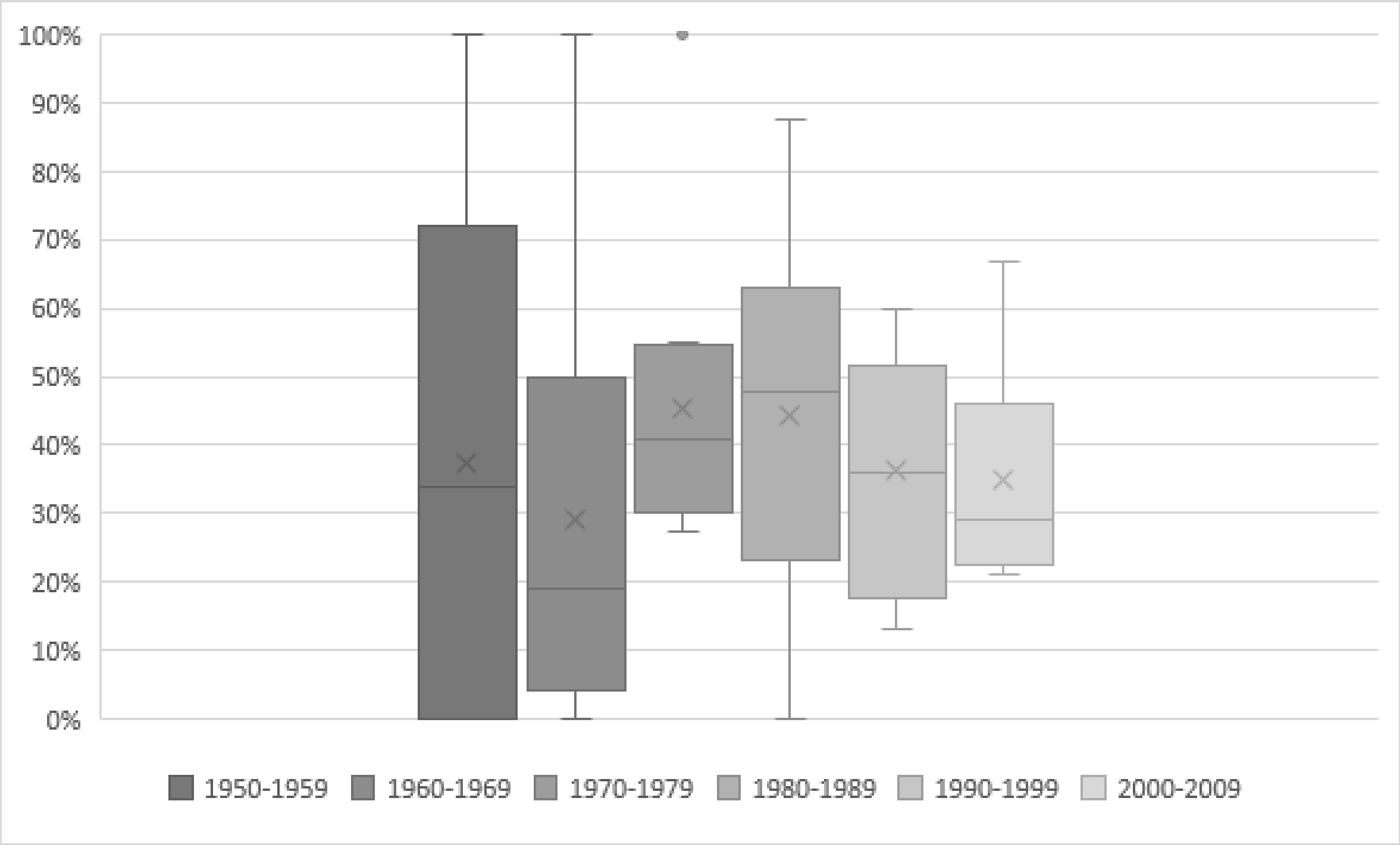

**Fig. 3** Percentage of Supporting citations over time (distribution over time). "×" = mean, "▬" = median, "o" = outlier.

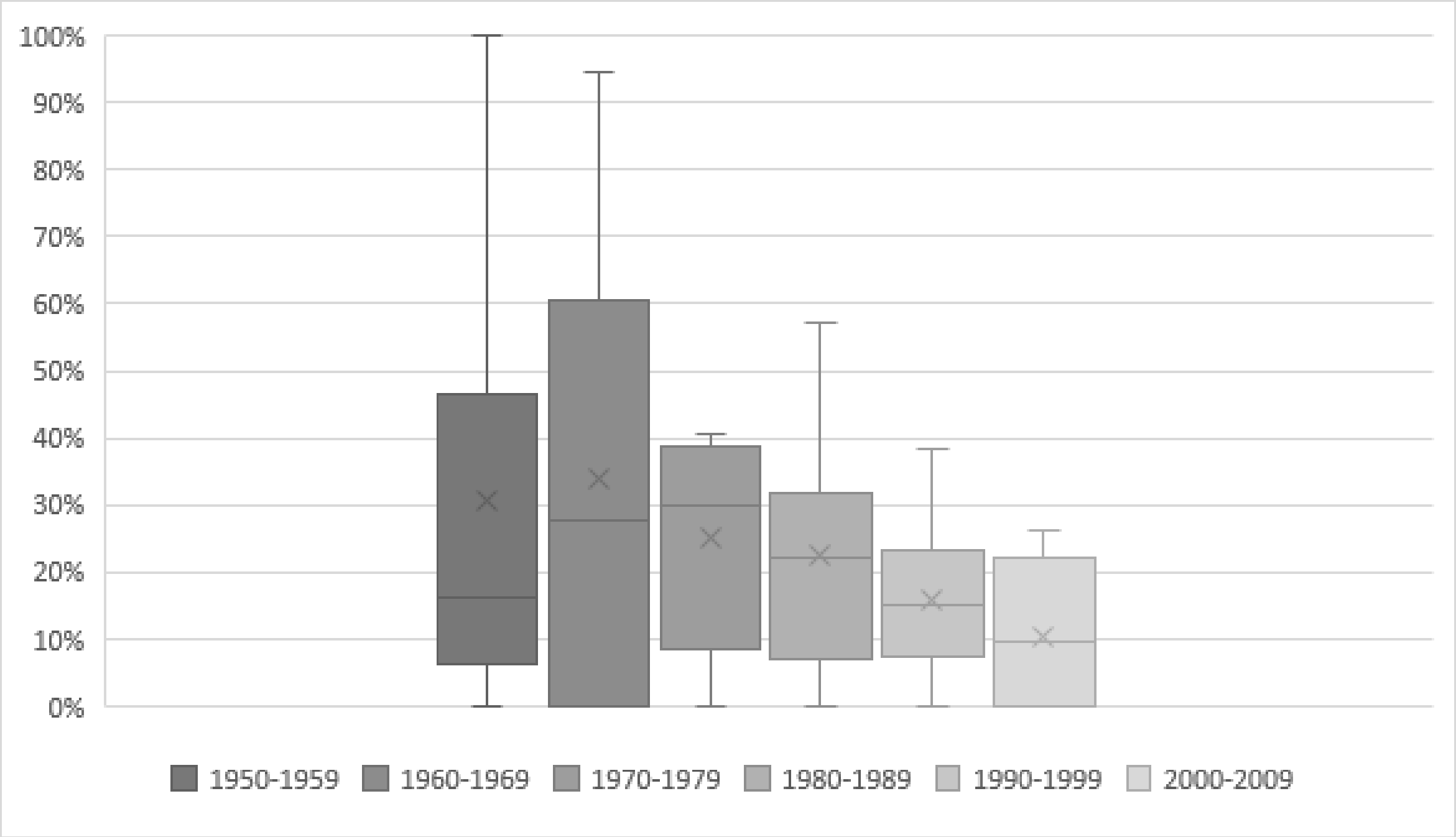

**Fig. 4** Percentage of Critical citations over time (distribution over time)

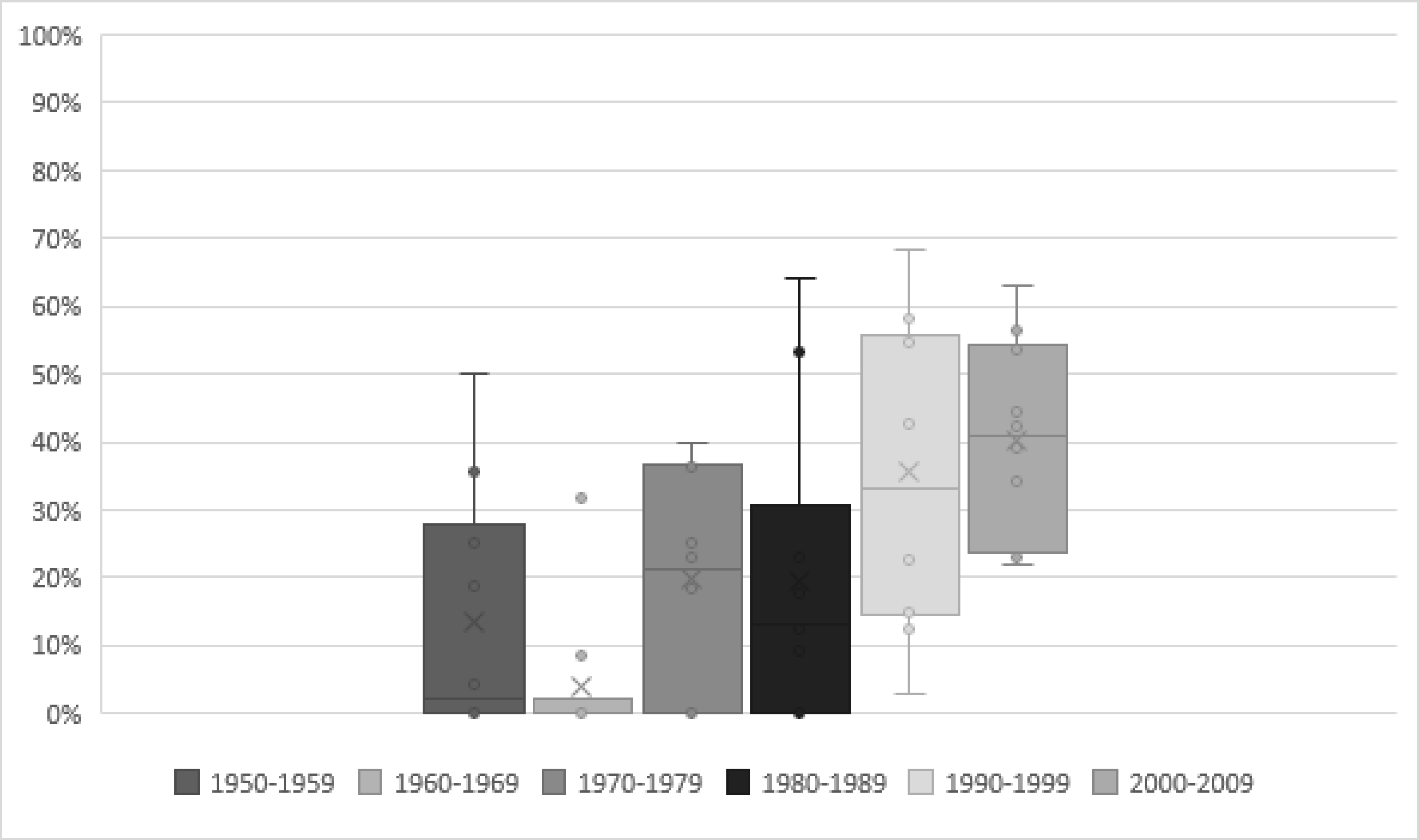

**Fig. 5** State of the art citations (distribution over time)

Fig. 6, 7 and 8 show the trend of mean and median for each of the three categories. Both mean and median are shown because of the non-normal distribution of the citations.

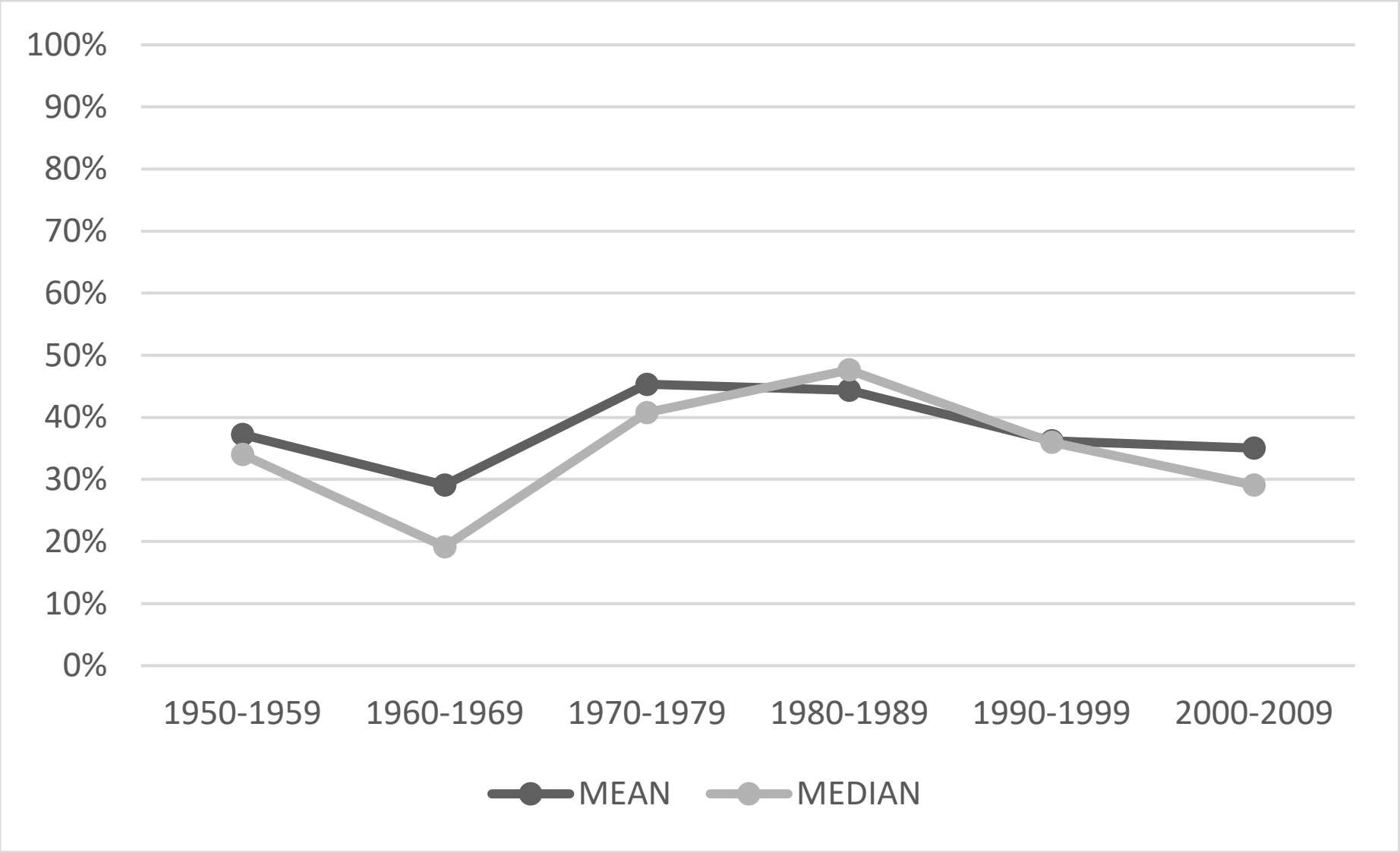

**Fig. 6** Percentage of Supporting citations (Mean and Median over time)

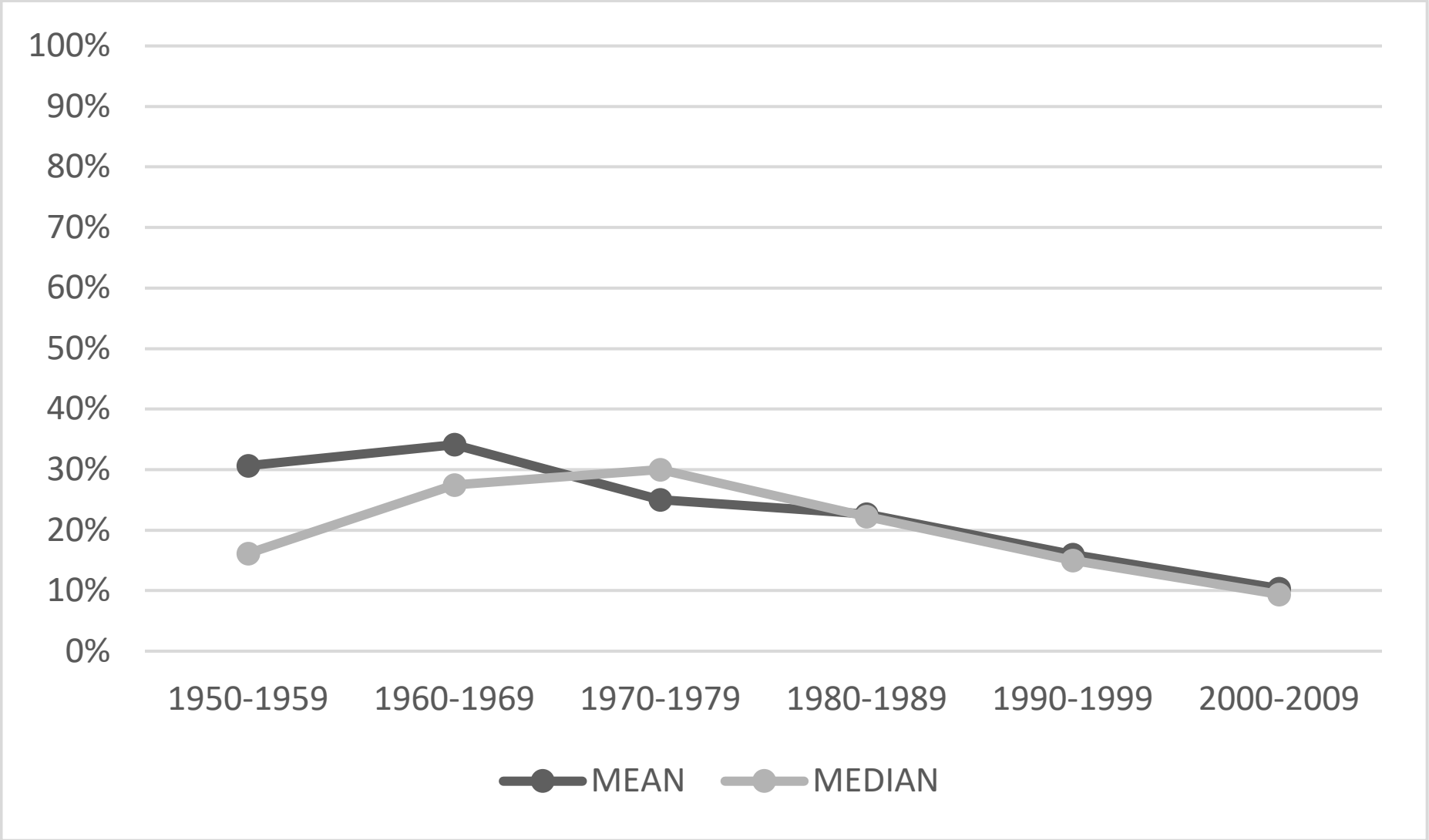

**Fig. 7** Percentage of Critical citations (Mean and Median over time)

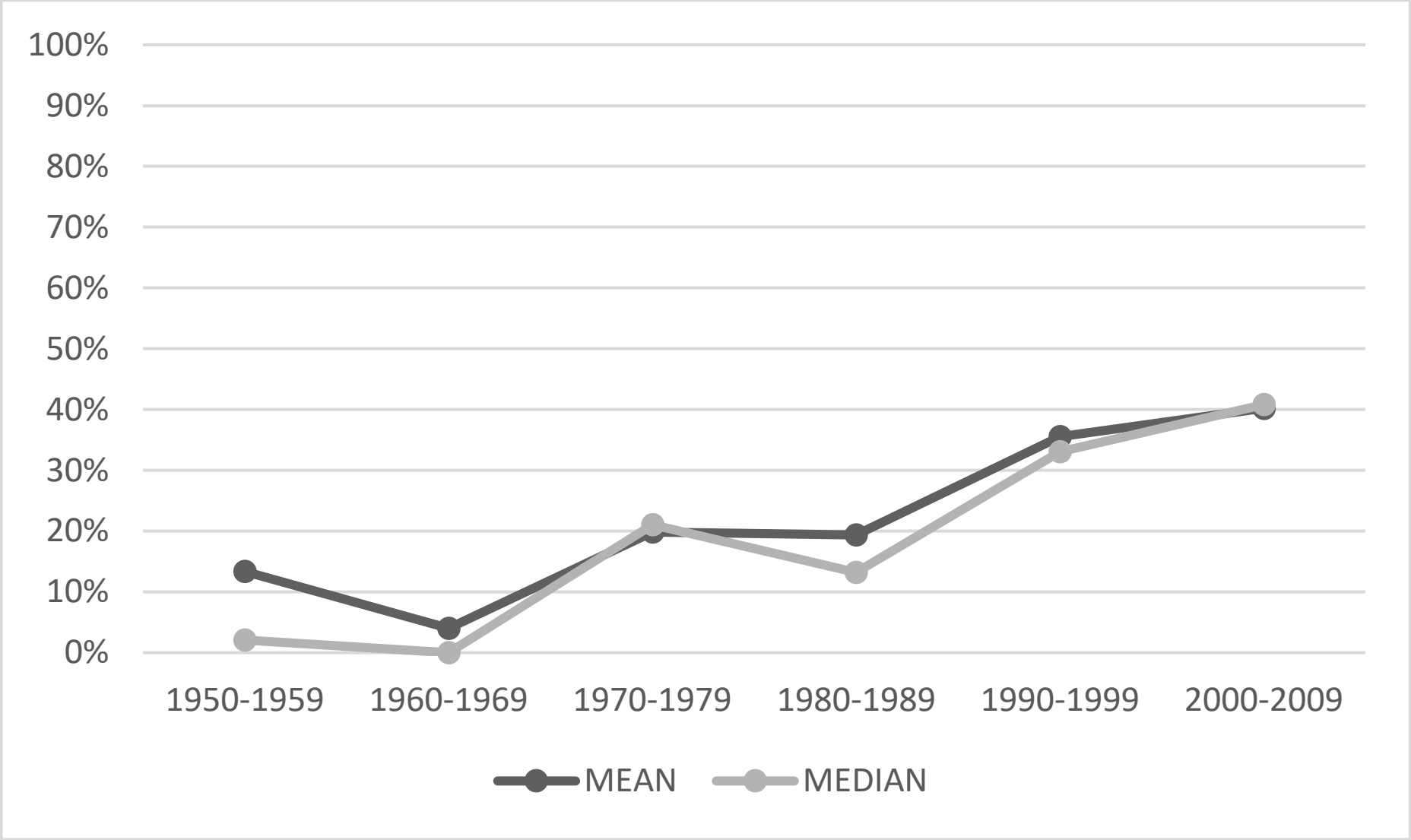

**Fig. 8** Percentage of State of the art citations (Mean and Median over time)

The percentage of State of the art citations clearly increases, especially in the last three decades. Their average triplicates from the 1950s to 2000s. The increase in the median values is even more evident (from 2.1% to 40.8%). From 1990 their minimum leaves the 0, becoming 2.9% in 1990s and 21.9% in 2000s. Furthermore, in the last decade, this category has become the most common type of citation, overtaking the Supporting citations.

Critical citations show an almost linear decreasing trend since the 1970s, with their mean decreasing from 30.7% in 1950s to 10.4% in 2010s. On the other hand, the trend of Supporting citations is more unstable and shows a slight decrease in the last decades.

Lastly, Fig. 9 depicts the evolution of the average percentage of Supplementary, Acknowledgement, and Documental citations. The only type showing a clear trend is the percentage of Supplementary citations, which reaches its maximum peak in the last decade.

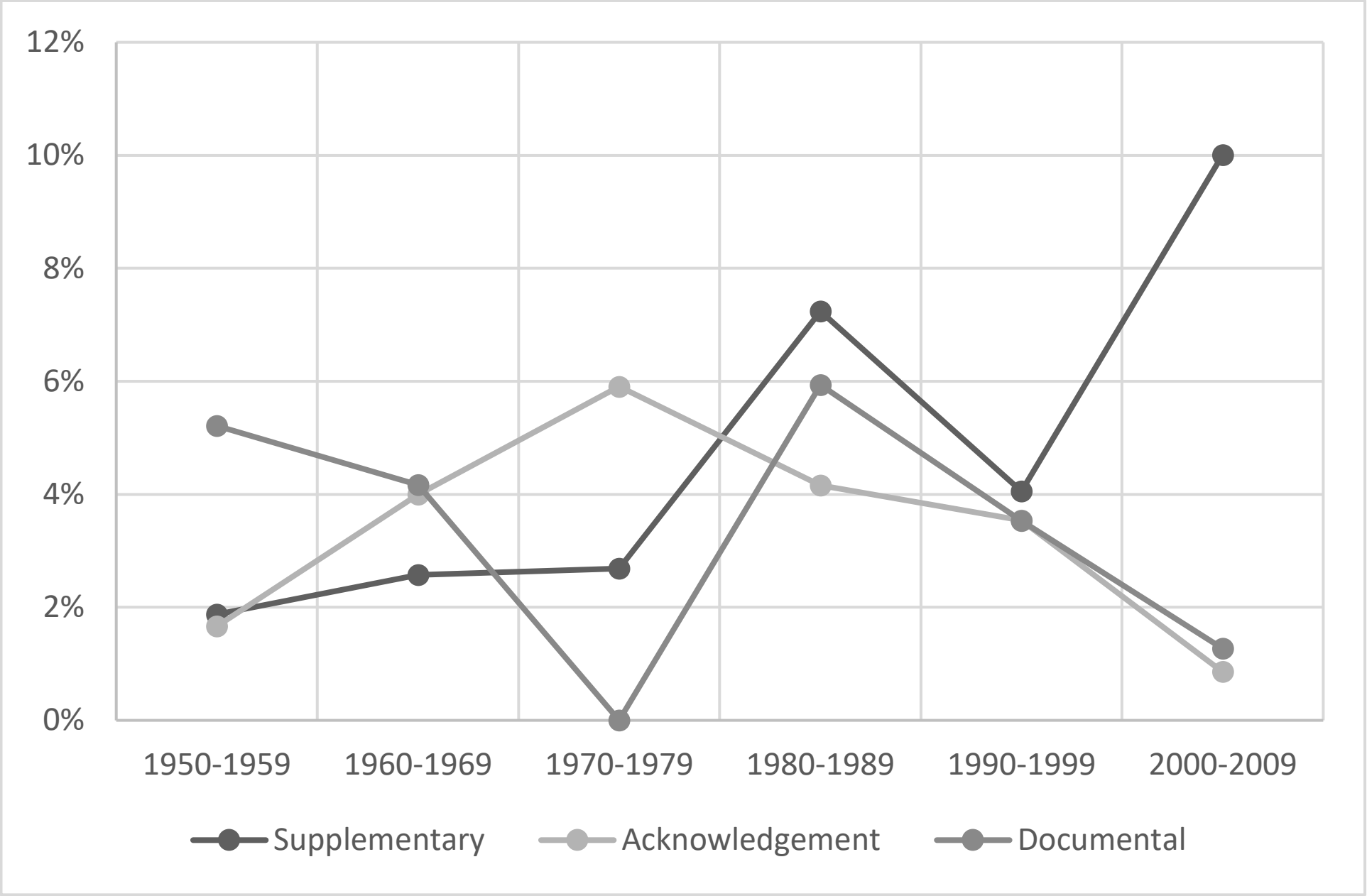

**Fig. 9** Percentage of Supplementary, Acknowledgment and Documental citations (Mean over time)

## 4 DISCUSSION

The first clear pattern in the data is the quasi-exponential growth in the average number of references per paper (Fig.1). This suggests an evolution in the citation behavior of analytic philosophers, who clearly started using more explicit references to the literature, especially in the last two decades. The interplay of three different factors may explain this pattern. Firstly, it is possible that editorial policies of AP journals changed in the last decades, encouraging authors to state explicit citations, rather than referring to implicit references. This may be in turn a consequence of the rise of the "citation culture" described by Wouters (1999). Secondly, it is possible that the emergence of the Internet has considerably simplified the literature search for analytic philosophers, in the same way as it helped scientists (Ucar et al. 2014). Indeed, there exist two databases dedicated to philosophical literature (*The Philosopher's Index* and the more recent *PhilPapers*) which are widely used by analytic philosophers for their literature searching. Thirdly, the exponential pattern may be explained by the growth of the disciplinary literature of AP, and as a sign of an *accumulation* process taking place.

Therefore, the first clear pattern in the data suggests that the answer to our research question (has AP started to *accumulate* during the second half of the twentieth century?) may be positive. However, the citations must be further analyzed to provide a clear answer to this research question, as their type and epistemological function should be taken into account.

The State of the Art category is particularly relevant to investigate the phenomenon of knowledge accumulation. The presence of this type of citations points out that the author of the citing paper regards her own work as contributing to an existing disciplinary body, to an on-going debate. State of the art citations are evidences of a collective work taking place, and the sign that a stock of knowledge is accumulating. Therefore, the increased percentage of this type of citations in the last 60 years (Fig. 5 and 8), and their widespread use in the 2000s, corroborates the idea that AP effectively underwent a process of accumulation. In addition to this, it is important to notice that in the 1990s and 2000s, we do not find any paper lacking this type of citations. In the 2000s, AP seems to have reached a mature stage of development, with an established literature shaping the intellectual production. This result is in accordance with evidence coming from co-citation analysis of AP. Buonomo and Petrovich (2018) report that after the turn of the century AP appears to be clearly divided into a set of sub-disciplines with relatively defined borders.

The decreasing trend in the percentage of Critical citations (Fig. 4 and 7, Fig. 2 even shows a decrease in the absolute number of Critical citations in the last decade) sheds light on another aspect of the normalization process, namely the decreasing weight of disagreement within the AP community.[17] This is also in line with Kuhn's theory of normal science and the low rate of negative citations found in the sciences (Catalini et al. 2015, Bertin et al. 2016, Cano 1989).

However, the changing trend in the percentage of Supporting citations, and their almost linear decrease in the last three decades (Fig. 3 and 6), implies that we have to be wary of a straightforward assimilation of AP to a classic Kuhnian normal science. It is important to remember that these citations are the only ones whose epistemological function is openly *constructive,* and the only ones stating an explicit *consensus* between the citing and the cited documents. State of the art citations, on the contrary, are cited without an open endorsement of the cited document. From an epistemological point of view, they are cited "neutrally" by the citing author. Their function is to build a background for the contribution, not to support some specific claim[18], so that they cannot be directly interpreted as fully positive citations. In AP, it is possible for an author to firstly cite several documents in order to review the topic's state of the art, and then to dismiss all their claims.

Given the lack of a well-defined increasing pattern in the occurrence of Supporting citations, the statement that AP is a Kuhnian Normal science must be honed. From the data, we can infer that AP is a specialty that has been consolidated in the last decades, without at the same time converging towards consensus. This apparently contradictory situation (accumulation without consensus) can be explained by the following hypothesis: AP underwent a process of *fragmentation* into several sub-disciplines, especially from the 1990s onwards. Under this hypothesis, the State of the art citations have the primary function of *identifying the sub-disciplinary area* to which the paper is meant to contribute. Once the sub-area is identified, analytic philosophers debate inside it in the classical philosophical fashion, i.e. citing both positively and negatively other documents. Within each sub-area, the consensus is lacking, as the decrease in the proportion of Supporting citations from the 1980s may indicate. However, the papers to contrast in order to advance a debate is clearly defined. Each sub-area has its own state-of-the-art. If this hypothesis was correct, the fragmentation process of AP would have begun in the 1980s, the decade in which the percentages of Supporting and State of the Art citations started following opposite trends (the first decreasing, the second increasing). This is not very far from traditional historical accounts of recent AP, which set in the middle 1970s the beginning of the fragmentation of the field (Tripodi 2015 : ch.4).

A kind of consensus emerges at the level of AP as a whole. Here there is a consensus about the *background structure of the field.* The accumulation process of AP therefore concerns the progressive stabilization of a sub-disciplinary organization, i.e. the increasing division of the epistemic labor into a set of specific sub-areas. This hypothesis provides a possible explanation for the patterns in the three most common types of citation.[19]

---

[17] The overall proportion of Critical citations (Tab. 2) is higher than the result reported by Cullars in his study of philosophy monographs, where 11.1% of citations were classified as critical (Cullars 1998 : 62). This is probably due to the fact that papers are more narrow-focused than books, and therefore devote more space to criticism and argumentation.

[18] Unless very loosely, by showing that the author is legitimated to contribute to the debate because she is up-to-date.

[19] As a Reviewer suggested, another hypothesis can be advanced to explain these results, namely that the fragmentation of the field is simply the consequence of the massive amount of information analytic philosophers have to confront with. As Quinn (1987) says: «Having limited time and energy at their disposal, individual philosophers have to focus rather narrowly to keep up with rapid developments in their areas of specialization» (111). A similar idea can be found in Marconi (2014)

The increasing trend in the percentage of Perfunctory citations (Fig. 9), especially in the last decade, may be a consequence of the general increase in the average number of references per paper. It could be argued that the longer the reference list, the higher is the probability that some of the cited works are not strictly relevant to the citing author's immediate concerns. With regard to the overall proportion of Perfunctory citations (Tab.2), this can be compared with results from other studies. Bornmann and Daniel (2008) report that the proportion of citations classified as "perfunctory" in previous studies ranged from about 10% to 50%. In the current study, Perfunctory citations reached an average of 10% only in the last decade. The overall low proportion may be due to the *discursive* nature of AP: analytic philosophers' research practice is based on a continuous debate with peers. Citations are mainly used as "moves in the epistemic game" (to locate in a specific stream of discussion – State of the Art citations – , to attack or defend a position – respectively Critical and Supporting citations). Using a Perfunctory citation in a paper is like doing a move without any effect, and this may be the reason why analytic philosophers are inclined to avoid this type of citations.[20]

Perhaps, the patterns in the last two categories are easier to explain. Acknowledgement citations were never a significative proportion of the sample (Tab. 2), and the decrease in their percentage is probably due to the simultaneous establishment of manuscript' sections specifically dedicated to acknowledgements (Cronin et al. 2003). Lastly, the low proportion of Documental citation (Tab. 2) is coherent with the theoretical focus of AP, whose main interest lies in philosophical theorizing rather than in historical reconstruction of past philosophers' thought.

## 5 CONCLUSIONS

In this study, qualitative citation context analysis was used to shed light on an epistemological question: how does knowledge accumulate in para-scientific areas? Analytic Philosophy was selected as an interesting example of para-science, and Kuhnian theory of normal science was chosen as a theoretical framework. The main result of the analysis is that knowledge in AP indeed started to accumulate but in a peculiar manner. Accumulation was not matched by a rising consensus within the AP community. This can be explained by the hypothesis that AP has undergone a process of fragmentation in sub-fields. In each sub-field, the discussion is characterized by high dialectics and low consensus rates that are typical of philosophical debates. However, a consensus arises at the level of the overall field, and it concerns the division of cognitive labor in AP, i.e. about the presence of specific sub-disciplinary literatures. Hence, it may be said that a sort of "soft paradigm" has taken over.[21]

In general, the results seem to confirm the claim, recently put forth by historians of philosophy, that AP underwent a process of normalization (i.e. assimilation to a Kuhnian normal science model) in the last decades. Citation context analysis allowed not only to empirically test this assertion, but also to study in detail patterns and specific features of the normalization process, proving to be a valuable tool for epistemological inquiries.

and Schwartz (1995), both linking the trend towards specialization to the growth of the literature available in AP. This hypothesis is worth considering because it brings into the picture an important factor influencing the citing behavior (i.e. the increase of the available literature). However, we believe this is not in contrast with the hypothesis advanced in this study (i.e. the normalization of AP in the form of the progressive delineation of a background structure of the field): normalization and specialization could be indeed two *mutually reinforcing factors* shaping contemporary AP. This interpretation is also consistent with Kuhn's late theory of science (see Kuhn 2000, where the philosophers explicitly linked specialization and normal science; see also Brad Wray 2011). A possible way to check the weight of the different factors could be, as a Reviewer suggested, to classify the contexts citing the 60 articles under consideration, and see if the proportions of categories change over time and across different sub-fields. We leave this for a future work.

[20] However, it is also possible that the understanding of "perfunctory/supplementary" citations used in this study is different from the one employed in previous ones.

[21] I thank an anonymous Reviewer for the expression "soft paradigm".

However, this study focused on the similarities between AP and Normal science at the level of *research practices*, as far as they can be captured by citation context analysis. Differences of *conceptual* status between the two fields – albeit clearly existing – were explicitly left outside the study. An important challenge for the future work is the integration of the conceptual dimension into studies of knowledge accumulation processes. This integration will allow the assessment of how conceptual differences between AP and science shape their accumulation patterns. Such study would require an even closer integration between philosophy of science and scientometrics than the one attempted in this paper. [22]

Further research is needed to connect the epistemological stance taken in this research (which focused on the accumulation of knowledge) to sociologically oriented theories of citations (which focus on practices of epistemic communities). Understanding which one of the most renowned citation theories better takes into account the accumulation process in para-scientific areas is certainly worth further investigation. Importantly, it would also clarify the validity of citation-based assessments of research performance in the context of para-scientific research.

[22] I thank an anonymous Reviewer for this suggestion.

# 7 APPENDIX 1: SAMPLE METADATA

| # | Author | PY | Title | Source | Vol | Is | First Page | End Page |
|---|---|---|---|---|---|---|---|---|
| | | | **1950-1959** | | | | | |
| 1 | GRICE, HP | 1957 | MEANING | Philos. Rev. | 66 | 3 | 377 | 388 |
| 2 | VENDLER, Z | 1957 | VERBS AND TIMES | Philos. Rev. | 66 | 2 | 143 | 160 |
| 3 | SMART, JJC | 1959 | SENSATIONS AND BRAIN PROCESSES | Philos. Rev. | 68 | 2 | 141 | 156 |
| 4 | RAWLS, J | 1958 | JUSTICE AS FAIRNESS | Philos. Rev. | 67 | 2 | 164 | 194 |
| 5 | QUINE, WV | 1956 | QUANTIFIERS AND PROPOSITIONAL ATTITUDES | J. Philos. | 53 | 5 | 177 | 187 |
| 6 | SEARLE, JR | 1958 | PROPER NAMES | Mind | 67 | 266 | 166 | 173 |
| 7 | FREGE, G | 1956 | THE THOUGHT - A LOGICAL INQUIRY | Mind | 65 | 259 | 289 | 311 |
| 8 | SIBLEY, F | 1959 | AESTHETIC CONCEPTS | Philos. Rev. | 68 | 4 | 421 | 450 |
| 9 | GRICE, HP; STRAWSON, PF | 1956 | IN DEFENSE OF A DOGMA | Philos. Rev. | 65 | 2 | 141 | 158 |
| 10 | DUMMETT, M | 1959 | WITTGENSTEIN PHILOSOPHY OF MATHEMATICS | Philos. Rev. | 68 | 3 | 324 | 348 |
| | | | **1960-1969** | | | | | |
| 1 | DONNELLAN, KS | 1966 | REFERENCE AND DEFINITE DESCRIPTIONS | Philos. Rev. | 75 | 3 | 281 | 304 |
| 2 | DAVIDSON, D | 1963 | ACTIONS, REASONS, AND CAUSES - SYMPOSIUM | J. Philos. | 60 | 23 | 685 | 700 |
| 3 | FRANKFURT, HG | 1969 | ALTERNATE POSSIBILITIES AND MORAL RESPONSIBILITY | J. Philos. | 66 | 23 | 829 | 839 |
| 4 | HARMAN, GH | 1965 | THE INFERENCE TO THE BEST EXPLANATION | Philos. Rev. | 74 | 1 | 88 | 95 |
| 5 | GRICE, HP | 1969 | UTTERERS MEANING AND INTENTIONS | Philos. Rev. | 78 | 2 | 147 | 177 |
| 6 | LEWIS, DK | 1968 | COUNTERPART THEORY AND QUANTIFIED MODAL LOGIC | J. Philos. | 65 | 5 | 113 | 126 |
| 7 | DANTO, A | 1964 | THE ARTWORLD | J. Philos. | 61 | 19 | 571 | 584 |
| 8 | BENACERRA | 1965 | WHAT NUMBERS COULD | Philos. | 74 | 1 | 47 | 73 |
| | F, P | | NOT BE | Rev. | | | | |
| 9 | STRAWSON, PF | 1964 | INTENTION AND CONVENTION IN SPEECH ACTS | Philos. Rev. | 73 | 4 | 439 | 460 |
| 10 | GEACH, PT | 1965 | ASSERTION | Philos. Rev. | 74 | 4 | 449 | 465 |
| | | | **1970-1979** | | | | | |
| 1 | NAGEL, T | 1974 | WHAT IS IT LIKE TO BE A BAT | Philos. Rev. | 83 | 4 | 435 | 450 |
| 2 | KRIPKE, S | 1975 | OUTLINE OF A THEORY OF TRUTH | J. Philos. | 72 | 19 | 690 | 716 |
| 3 | PERRY, J | 1979 | PROBLEM OF THE ESSENTIAL INDEXICAL | Nous | 13 | 1 | 3 | 21 |
| 4 | CUMMINS, R | 1975 | FUNCTIONAL-ANALYSIS | J. Philos. | 72 | 20 | 741 | 765 |
| 5 | LEWIS, D | 1979 | ATTITUDES DE-DICTO AND DE-SE | Philos. Rev. | 88 | 4 | 513 | 543 |
| 6 | GOLDMAN, AI | 1976 | DISCRIMINATION AND PERCEPTUAL KNOWLEDGE | J. Philos. | 73 | 20 | 771 | 791 |
| 7 | LEWIS, D | 1979 | COUNTERFACTUAL DEPENDENCE AND TIMES ARROW | Nous | 13 | 4 | 455 | 476 |
| 8 | DRETSKE, FI | 1970 | EPISTEMIC OPERATORS | J. Philos. | 67 | 24 | 1007 | 1023 |
| 9 | LEWIS, D | 1976 | PROBABILITIES OF CONDITIONALS AND CONDITIONAL PROBABILITIES | Philos. Rev. | 85 | 3 | 297 | 315 |
| 10 | WRIGHT, L | 1973 | FUNCTIONS | Philos. Rev. | 82 | 2 | 139 | 168 |
| | | | **1980-1989** | | | | | |
| 1 | CHURCHLAND, PM | 1981 | ELIMINATIVE MATERIALISM AND THE PROPOSITIONAL ATTITUDES | J. Philos. | 78 | 2 | 67 | 90 |
| 2 | RAWLS, J | 1980 | RATIONAL AND FULL AUTONOMY | J. Philos. | 77 | 9 | 515 | 535 |
| 3 | JACKSON, F | 1986 | WHAT MARY DIDNT KNOW + KNOWLEDGE ARGUMENT AGAINST PHYSICALISM | J. Philos. | 83 | 5 | 291 | 295 |
| 4 | RAILTON, P | 1986 | MORAL REALISM + A FORM OF ETHICAL NATURALISM | Philos. Rev. | 95 | 2 | 163 | 207 |
| 5 | WOLF, S | 1982 | MORAL SAINTS + IMPLICATIONS FOR MORAL-PHILOSOPHY | J. Philos. | 79 | 8 | 419 | 439 |
| 6 | BURGE, T | 1986 | INDIVIDUALISM AND PSYCHOLOGY | Philos. Rev. | 95 | 1 | 3 | 45 |
| 7 | KIM, J | 1984 | CONCEPTS OF | Philos. | 45 | 2 | 153 | 176 |
| | | | SUPERVENIENCE | Phenomenol. Res. | | | | |
| 8 | BURGE, T | 1988 | INDIVIDUALISM AND SELF-KNOWLEDGE | J. Philos. | 85 | 11 | 649 | 663 |
| 9 | MCGINN, C | 1989 | CAN WE SOLVE THE MIND BODY PROBLEM | Mind | 98 | 391 | 349 | 366 |
| 10 | BOGEN, J; WOODWARD, J | 1988 | SAVING THE PHENOMENA | Philos. Rev. | 97 | 3 | 303 | 352 |
| | | | **1990-1999** | | | | | |
| 1 | DEROSE, K | 1995 | SOLVING THE SKEPTICAL PROBLEM | Philos. Rev. | 104 | 1 | 1 | 52 |
| 2 | BRATMAN, ME | 1992 | SHARED COOPERATIVE ACTIVITY | Philos. Rev. | 101 | 2 | 327 | 340 |
| 3 | BURGE, T | 1993 | CONTENT PRESERVATION | Philos. Rev. | 102 | 4 | 457 | 488 |
| 4 | YABLO, S | 1992 | MENTAL CAUSATION | Philos. Rev. | 101 | 2 | 245 | 280 |
| 5 | DENNETT, DC | 1991 | REAL PATTERNS | J. Philos. | 88 | 1 | 27 | 51 |
| 6 | EDGINGTON, D | 1995 | ON CONDITIONALS | Mind | 104 | 414 | 235 | 329 |
| 7 | DAVIDSON, D | 1990 | THE STRUCTURE AND CONTENT OF TRUTH | J. Philos. | 87 | 6 | 279 | 328 |
| 8 | LEWIS, D | 1994 | CHANCE AND CREDENCE - HUMEAN SUPERVENIENCE DEBUGGED | Mind | 103 | 412 | 473 | 490 |
| 9 | DEROSE, K | 1992 | CONTEXTUALISM AND KNOWLEDGE ATTRIBUTIONS | Philos. Phenomenol. Res. | 52 | 4 | 913 | 929 |
| 10 | GRIFFITHS, PE; GRAY, RD | 1994 | DEVELOPMENTAL SYSTEMS AND EVOLUTIONARY EXPLANATION | J. Philos. | 91 | 6 | 277 | 304 |
| | | | **2000-2009** | | | | | |
| 1 | Pryor, J | 2000 | The skeptic and the dogmatist | Nous | 34 | 4 | 517 | 549 |
| 2 | Stanley, J; Williamson, T | 2001 | Knowing how | J. Philos. | 98 | 8 | 411 | 444 |
| 3 | Lewis, D | 2000 | Causation as influence | J. Philos. | 97 | 4 | 182 | 197 |
| 4 | Kolodny, N | 2005 | Why be rational? | Mind | 114 | 455 | 509 | 563 |
| 5 | Elga, A | 2007 | Reflection and disagreement | Nous | 41 | 3 | 478 | 502 |
| 6 | Nichols, S; Knobe, J | 2007 | Moral responsibility and determinism: The cognitive | Nous | 41 | 4 | 663 | 685 |
| | | | science of folk intuitions | | | | | |
| 7 | Chalmers, DJ; Jackson, F | 2001 | Conceptual analysis and reductive explanation | Philos. Rev. | 110 | 3 | 315 | 360 |
| 8 | Rupert, RD | 2004 | Challenges to the hypothesis of extended cognition | J. Philos. | 101 | 8 | 389 | 428 |
| 9 | Christensen, D | 2007 | Epistemology of Disagreement: The Good News | Philos. Rev. | 116 | 2 | 187 | 217 |
| 10 | Byrne, A | 2001 | Intentionalism defended | Philos. Rev. | 110 | 2 | 199 | 240 |

## 8 APPENDIX 2: EXAMPLES OF CATEGORIES OF CITATION

In this Appendix, examples of categories of citations (see Tab.1) are reported. However, one should keep in mind that generally it is necessary to consider the *whole* content of the paper in order to classify plausibly the function of the citations. Very often, in the case of AP the "context" of the citation turns out to be the entire paper.

*State of the art citations:*

> "Almost all the extensive recent literature seeking alternatives to the orthodox approach – I would mention especially the writings of Bas an Fraassen and Robert L. Martin – agrees on a single basic idea…" [Kripke 1975 : 698]
>
> "For the classic discussion of these problems, see [12]" [Perry 1979 : 21]
>
> "A survey of the recent philosophical literature on the nature of functional analysis and explanation, beginning with the classic essays of Hempel in 1959 and Nagel in 1961, reveals that… [note]" [Cummins 1975 : 741]
>
> "Some philosophers have claimed that people have incompatibilist intuitions (e.g. Kane 1999, 218; Strawson 1986, 30; Vargas 2006); others have challenged this claim and suggested that people's intuitions actually fit with compatibilism (Nahmias et al. 2015)" [Nichols and Knobe 2007 : 663]

*Supporting citations:*

> "I follow Arthur Smullyan's treatment of scope ambiguity in modal sentences, given in 'Modality and Description', *Journal of Symbolic Logic*, XIII, 1 (March 1948): 31-37, as qualified by Wilson's objection, in *The Concept of Language*…" [Lewis 1968 : 120]
>
> "The influence of H. P. Grice's 'Meaning', *The Philosophical Review*, LXVI (1957): 377-388 will be evident here" [Davidson 1992 : 311]

*Supplementary citations:*

> "For my former view, see the treatment of preemption in 'Postscript E to 'Causation", in my *Philosophical Papers*, Volume II (New York: Oxford, 1986), pp. 193-212." [Lewis 2000 : 1983, in footnote]

*Acknowledgement citations:*

> "In thinking about the problem of essential indexical, I have been greatly helped by the writings of Hector-Neri Castaneda on indexicality and related topics. Castaneda focused attention on these problems, and made many of the points made here, in [1], [2] and [3]" [Perry 1979 : 21]

*Critical citations:*

> "The difficulty one gets into by a mechanical application of the theory of games to moral philosophy can be brought out by considering among several possible examples, R. B. Braithwaite's study, *Theory of Games as a Tool for the Moral Philosopher* (Cambridge 1955) […] Braithwaite's use of the theory of games, insofar as it is intended to analyze the concept of fairness, is, I think, mistaken" [Rawls 1958 : 176-177]

> "W. V. Quine, for one, explicitly denies that anything need to be done other than provide a progression to serve as the numbers. In *Word and Object* (London, 1960, pp. 262-263, he states […] I would disagree" [Benacerraf 1965 : 51]

*Documental citations:*

> "While the assimilation is implicit in Bentham's and Sidgwick's moral theory, explicit statements of it as applied to justice are relatively rare. One clear instance in *The Principles of Morals and Legislation* occurs in ch. X, footnote 2 to section XL: […]" [Rawls 1958 : 184]

## 9 APPENDIX 3: RAW DATA

In the following Table, percentage of citation categories, divided by paper and decade, are reported. Papers' profiles are available on-line in Petrovich (2008).

| Supporting citations | | | | | | |
|---|---|---|---|---|---|---|
| # Paper | 1950-1959 | 1960-1969 | 1970-1979 | 1980-1989 | 1990-1999 | 2000-2009 |
| 1 | 0.00% | 8.33% | 42.31% | 52.94% | 18.84% | 21.01% |
| 2 | 66.67% | 5.56% | 54.55% | 20.00% | 31.37% | 31.25% |
| 3 | 87.50% | 0.00% | 30.77% | 42.86% | 46.00% | 61.54% |
| 4 | 32.29% | 100.00% | 33.33% | 52.38% | 36.05% | 22.62% |
| 5 | 100.00% | 50.00% | 39.13% | 0.00% | 56.25% | 32.86% |
| 6 | 50.00% | 38.64% | 28.13% | 66.67% | 14.22% | 21.74% |
| 7 | 0.00% | 0.00% | 42.86% | 24.19% | 35.96% | 25.00% |
| 8 | 35.71% | 30.00% | 100.00% | 61.54% | 50.00% | 40.74% |
| 9 | 0.00% | 50.00% | 55.00% | 87.50% | 13.16% | 66.67% |
| 10 | 0.00% | 8.33% | 27.27% | 35.23% | 60.00% | 26.86% |
| **MEAN** | **37.22%** | **29.09%** | **45.33%** | **44.33%** | **36.19%** | **35.03%** |
| **MEDIAN** | **34.00%** | **19.17%** | **40.72%** | **47.62%** | **36.01%** | **29.06%** |
| **MAX** | **100.00%** | **100.00%** | **100.00%** | **87.50%** | **60.00%** | **66.67%** |
| **MIN** | **0.00%** | **0.00%** | **27.27%** | **0.00%** | **13.16%** | **21.01%** |
| **Critical citations** | | | | | | |
| # Paper | 1950-1959 | 1960-1969 | 1970-1979 | 1980-1989 | 1990-1999 | 2000-2009 |
| 1 | 100.00% | 50.00% | 38.46% | 47.06% | 23.91% | 2.17% |
| 2 | 16.67% | 94.44% | 18.18% | 26.67% | 7.84% | 25.00% |
| 3 | 8.33% | 0.00% | 7.69% | 57.14% | 20.67% | 0.00% |
| 4 | 15.63% | 0.00% | 26.67% | 14.29% | 0.00% | 26.19% |
| 5 | 0.00% | 50.00% | 8.70% | 7.14% | 16.67% | 0.00% |
| 6 | 25.00% | 0.00% | 40.63% | 22.22% | 5.60% | 0.00% |
| 7 | 0.00% | 0.00% | 33.33% | 6.45% | 22.81% | 21.15% |
| 8 | 28.57% | 30.00% | 0.00% | 23.08% | 38.24% | 9.26% |
| 9 | 100.00% | 25.00% | 40.00% | 0.00% | 10.53% | 9.52% |
| 10 | 12.50% | 91.67% | 36.36% | 22.35% | 13.33% | 10.36% |
| **MEAN** | **30.67%** | **34.11%** | **25.00%** | **22.64%** | **15.96%** | **10.37%** |
| **MEDIAN** | **16.15%** | **27.50%** | **30.00%** | **22.29%** | **15.00%** | **9.39%** |
| **MAX** | **100.00%** | **94.44%** | **40.63%** | **57.14%** | **38.24%** | **26.19%** |
| **MIN** | **0.00%** | **0.00%** | **0.00%** | **0.00%** | **0.00%** | **0.00%** |
| **State of the art citations** | | | | | | |
| # Paper | 1950-1959 | 1960-1969 | 1970-1979 | 1980-1989 | 1990-1999 | 2000-2009 |
| 1 | 0.00% | 8.33% | 0.00% | 0.00% | 42.75% | 63.04% |
| 2 | 0.00% | 0.00% | 18.18% | 9.17% | 54.90% | 21.88% |
| 3 | 4.17% | 0.00% | 23.08% | 0.00% | 22.67% | 23.08% |
| 4 | 18.75% | 0.00% | 40.00% | 0.00% | 54.65% | 39.29% |
| 5 | 0.00% | 0.00% | 36.96% | 64.29% | 12.50% | 44.29% |
| 6 | 25.00% | 31.82% | 25.00% | 13.89% | 58.19% | 56.52% |
| 7 | 0.00% | 0.00% | 19.05% | 53.23% | 14.91% | 42.31% |
| 8 | 35.71% | 0.00% | 0.00% | 23.08% | 2.94% | 34.26% |
| 9 | 0.00% | 0.00% | 0.00% | 12.50% | 68.42% | 23.81% |
| **10** | 50.00% | 0.00% | 36.36% | 17.80% | 23.33% | 53.56% |
| **MEAN** | **13.36%** | **4.02%** | **19.86%** | **19.40%** | **35.53%** | **40.20%** |
| **MEDIAN** | **2.09%** | **0.00%** | **21.06%** | **13.20%** | **33.04%** | **40.80%** |
| **MAX** | **50.00%** | **31.82%** | **40.00%** | **64.29%** | **68.42%** | **63.04%** |
| **MIN** | **0.00%** | **0.00%** | **0.00%** | **0.00%** | **2.94%** | **21.88%** |
| **Supplementary/Perfunctory citations** | | | | | | |
| **# Paper** | **1950-1959** | **1960-1969** | **1970-1979** | **1980-1989** | **1990-1999** | **2000-2009** |
| **1** | 0.00% | 16.67% | 11.54% | 0.00% | 8.70% | 9.42% |
| **2** | 0.00% | 0.00% | 0.00% | 20.00% | 0.00% | 12.50% |
| **3** | 0.00% | 0.00% | 0.00% | 0.00% | 4.00% | 15.38% |
| **4** | 18.75% | 0.00% | 0.00% | 4.76% | 4.65% | 7.14% |
| **5** | 0.00% | 0.00% | 4.35% | 14.29% | 4.17% | 14.29% |
| **6** | 0.00% | 9.09% | 6.25% | 5.56% | 3.88% | 15.22% |
| **7** | 0.00% | 0.00% | 4.76% | 0.00% | 7.02% | 11.54% |
| **8** | 0.00% | 0.00% | 0.00% | 7.69% | 5.88% | 9.26% |
| **9** | 0.00% | 0.00% | 0.00% | 0.00% | 0.00% | 0.00% |
| **10** | 0.00% | 0.00% | 0.00% | 20.08% | 2.22% | 5.34% |
| **MEAN** | **1.88%** | **2.58%** | **2.69%** | **7.24%** | **4.05%** | **10.01%** |
| **MEDIAN** | **0.00%** | **0.00%** | **0.00%** | **5.16%** | **4.08%** | **10.48%** |
| **MAX** | **18.75%** | **16.67%** | **11.54%** | **20.08%** | **8.70%** | **15.38%** |
| **MIN** | **0.00%** | **0.00%** | **0.00%** | **0.00%** | **0.00%** | **0.00%** |
| **Acknowledgment citations** | | | | | | |
| **# Paper** | **1950-1959** | **1960-1969** | **1970-1979** | **1980-1989** | **1990-1999** | **2000-2009** |
| **1** | 0.00% | 0.00% | 0.00% | 0.00% | 5.80% | 0.00% |
| **2** | 16.67% | 0.00% | 9.09% | 5.00% | 5.88% | 0.00% |
| **3** | 0.00% | 0.00% | 38.46% | 0.00% | 0.00% | 0.00% |
| **4** | 0.00% | 0.00% | 0.00% | 19.05% | 0.00% | 0.00% |
| **5** | 0.00% | 0.00% | 6.52% | 7.14% | 8.33% | 8.57% |
| **6** | 0.00% | 0.00% | 0.00% | 5.56% | 3.45% | 0.00% |
| **7** | 0.00% | 0.00% | 0.00% | 4.84% | 0.00% | 0.00% |
| **8** | 0.00% | 40.00% | 0.00% | 0.00% | 2.94% | 0.00% |
| **9** | 0.00% | 0.00% | 5.00% | 0.00% | 7.89% | 0.00% |
| **10** | 0.00% | 0.00% | 0.00% | 0.00% | 1.11% | 0.00% |
| **MEAN** | **1.67%** | **4.00%** | **5.91%** | **4.16%** | **3.54%** | **0.86%** |
| **MEDIAN** | **0.00%** | **0.00%** | **0.00%** | **2.42%** | **3.19%** | **0.00%** |
| **MAX** | **16.67%** | **40.00%** | **38.46%** | **19.05%** | **8.33%** | **8.57%** |
| **MIN** | **0.00%** | **0.00%** | **0.00%** | **0.00%** | **0.00%** | **0.00%** |
| **Documental citations** | | | | | | |
| **# Paper** | **1950-1959** | **1960-1969** | **1970-1979** | **1980-1989** | **1990-1999** | **2000-2009** |
| **1** | 0.00% | 16.67% | 0.00% | 0.00% | 0.00% | 0.00% |
| **2** | 0.00% | 0.00% | 0.00% | 19.17% | 0.00% | 6.25% |
| **3** | 0.00% | 0.00% | 0.00% | 0.00% | 6.67% | 0.00% |
| **4** | 14.58% | 0.00% | 0.00% | 9.52% | 4.65% | 0.00% |
| **5** | 0.00% | 0.00% | 0.00% | 7.14% | 2.08% | 0.00% |
| **6** | 0.00% | 0.00% | 0.00% | 0.00% | 2.59% | 0.00% |
| **7** | 0.00% | 0.00% | 0.00% | 11.29% | 19.30% | 0.00% |
| **8** | 0.00% | 0.00% | 0.00% | 7.69% | 0.00% | 6.48% |
| **9** | 0.00% | 25.00% | 0.00% | 0.00% | 0.00% | 0.00% |
| **10** | 37.50% | 0.00% | 0.00% | 4.55% | 0.00% | 0.00% |
| **MEAN** | **5.21%** | **4.17%** | **0.00%** | **5.94%** | **3.53%** | **1.27%** |
| **MEDIAN** | **0.00%** | **0.00%** | **0.00%** | **5.84%** | **1.04%** | **0.00%** |
| **MAX** | **37.50%** | **25.00%** | **0.00%** | **19.17%** | **19.30%** | **6.48%** |
| **MIN** | **0.00%** | **0.00%** | **0.00%** | **0.00%** | **0.00%** | **0.00%** |